\documentclass[aps,prl,twocolumn,showpacs,superscriptaddress,groupedaddress,reprint]{revtex4-1}

\usepackage{subfigure}
\usepackage{wrapfig}
\usepackage{textcomp}
\usepackage{graphicx}
\usepackage{amssymb}
\usepackage{amsmath}
\usepackage{bm}
\usepackage{amsmath, amsthm, amssymb}
\usepackage{hyperref}
\usepackage{color}

\usepackage{amsfonts}
\usepackage{dcolumn}
\usepackage{float}
\usepackage{bm}

\def\beq{\begin{equation}}
\def\eeq{\end{equation}}
\def\bea{\begin{eqnarray}}
\def\eea{\end{eqnarray}}

\begin{document}

\title{Coupled Wavepackets for Non-Adiabatic Molecular Dynamics: A Generalization of Gaussian Wavepacket Dynamics to Multiple Potential Energy Surfaces}

\author{Alexander White, Sergei Tretiak and Dmitry Mozyrsky}

\affiliation{Theoretical Division, Los Alamos National Laboratory, Los Alamos, New Mexico 87545, USA}

 \date{\today}

\begin{abstract}

	Accurate simulation of the non-adiabatic dynamics of molecules in excited electronic states is key to understanding molecular photo-physical processes. Here we present a novel method, based on a semiclassical approximation, that is as efficient as the commonly used mean field Ehrenfest or \emph{ad hoc} Surface Hopping methods and properly accounts for interference and decoherence effects.  This novel method is an extension of Heller's Thawed Gaussian wavepacket dynamics that includes coupling between potential energy surfaces. The accuracy of the method can be systematically improved. 

\end{abstract}

\maketitle

First principles-based molecular dynamics (MD) is becoming an important tool for understanding properties of complex molecular systems.\cite{ABMD,Tuckerman02,Car85} Unfortunately, the cost of exact dynamics, by direct calculation of the time-dependent Schr{\"o}dinger equation (TDSE), scales exponentially with the dimensionality (\emph{i.e.} number of atoms) of the system.\cite{Feit82,Kosloff84,Park86,Kosloff94,Neuhauser94,Guo14,Guo15} Thus, for large systems one often approximates that the nuclei of a molecule propagate via classical equations of motion and calculates forces (due to Coulombic interaction) via quantum chemistry methods. In doing so one typically relies on the \emph{Born-Oppenheimer}  approximation, where electrons remain in the same electronic quantum state $|n ({\bf x})\rangle$ with energy, $E^{(n)}(\bf{x})$, that parametrically depends on nuclear coordinates, ${\bf x}=(x_1, ..., x_N)^{T}$).\cite{Born} Thus, the nuclei propagate on a single potential energy surface (PES). For molecules in the ground electronic state, and at low temperatures, this situation often holds due to a sufficiently wide gap between the PES of the ground and excited electronic states. However, for certain nuclear configurations, common when the molecule is in an excited electronic state due to absorption of energy (\emph{e.g.} a photon), the gaps can become small or even vanish. In these regions, where the nuclear-electronic coupling is the same order as the energy gap, non-adiabatic behavior is expected.\cite{Massey}  This creates a superposition of electronic states, with different forces acting on the nuclei.
\begin{figure}[t]
\vspace{0.25 cm}	
\includegraphics[width=7 cm]{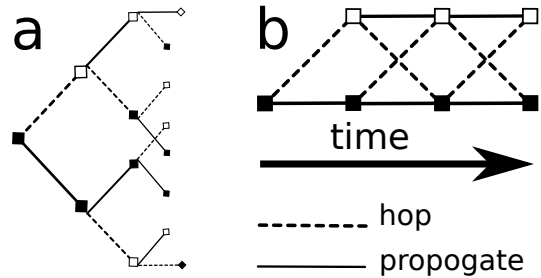}
\caption{\label{drawing} a- Branching Tree Solution to time dependent Schr{\" o}dinger equation (sampled by Monte-Carlo). b- Coupled wavepackets for non-adiabatic molecular dynamics (CW-NAMD) approximation to the Branching Tree. c- CW-NAMD approximation with coarse branching.} 
\end{figure}

 Since the full TDSE is numerically intractable for high dimensions, approximations for the non-adiabatic molecular dynamics (NAMD) must be made. The simplest approximation is to average over the electronic degree of freedom (DOF), a {mean-filed} approximation, to determine the force on the nuclei.\cite{Ehrenfest27,Nitzan85} This is known as the {Ehrenfest} approximation. Like any mean-field approximation, it breaks down when there is non-negligible correlation between the dynamical DOF (the nuclear) and the averaged DOF (the electronic), \emph{i.e.} if the components of the nuclear wavefunction separate depending on which PES they propogate. In an attempt to correct for this problem, while maintaining efficiency and simplicity, Tully proposed the surface hopping method,\cite{Tully71} most commonly used with the fewest-switching procedure (FSSH).\cite{Tully90}  In this method a swarm of classical trajectories propagate on an initial PES, with a finite probability to hop to a coupled PES in regions of non-adiabatic coupling. This method is \emph{ad-hoc}, and is only strictly accurate in the same limit as the Ehrenfest approximation.\cite{Subotnik10, Subotnik11} This  incomplete treatment of the nuclear-electron correlation has two well known symptoms: the {interference} problem, where the incorrect phase of the nuclear wavefunction leads to incorrect levels of constructive/deconstructive interference,  and the {decoherence} problem, where the separation of the nuclear wavefunction is improperly accounted for. Both problems were pointed out by Tully himself.\cite{Tully90} These two methods, Ehrenfest and FSSH, are by far the most commonly used in the simulation of NAMD.\cite{Kilina09,Kenichiro12,Nelson14,Akimov14,Wang15,Choi15,Goyal15,Petit15,Du15,Wenjun15,Wenjie15,Dou15,Galperin15} Many attempts have been made to improve upon the basic foundation of these two methods.\cite{Bittner95,Zhu04,Bedard05,Subotnik10, Subotnik11,Nelson13, Squash, Wang14, Sifain15, FSLS, Trivedi15} More sophisticated and accurate mixed quantum-classical and semi-classical methods, which are not \emph{ad-hoc}, are typically applied only in small, or reduced, systems due to inefficiency and/or complexity.\cite{Meyera79,Thoss99,Kapral06,Chen06,Miller09,Herman05,Rassolov05,Zamstein12,Gorshkov2013,Menzeleev14,White14,Makri15,Gross15,Pfalzgraff15,Martens15} 

An ideal NAMD method would have certain properties. It must (1) be based on localized dynamics, \emph{i.e.} based on real-space trajectories, (2) use only local parameters easily calculated from common electronic structure methods, \emph{i.e.} PES and electronic wavefunction,  (3) require no empirical or \emph{ad-hoc} treatments, (4) include proper treatment of electron-nuclear coupling, (5) be at least as efficient as surface hopping, and (6) be systematically improvable.  To build a new NAMD method, which satisfies the first and second conditions, one must start from a sound foundation for these real-spaced trajectories. The closest analog to a classical particle, and thus real local trajectory, for a quantum system is a localized wave packet, or superposition of wave packets. \cite{PQM}

The use of complex multi-dimensional Gaussian wave packets (GWP):
\begin{align}
\label{gauss}
g({\bf x}; {\bf x}_0, {\bf p}_0, {\hat \alpha}_0) = e^{\frac{i}{\hbar}  [\gamma_0 + {\bf p}_0^T({\bf x}-{\bf x}_0) + ({\bf x}-{\bf x}_0)^T{\hat \alpha}_0({\bf x}-{\bf x}_0)]} ,
\end{align}
as approximations, or basis sets, for nuclei wavefunctions is well studied for semiclassical dynamics on a single potential energy surfaces.\cite{Heller75,Heller81,Huber87,Coalson90,Pattanayak94,Herman84,Kluk86,Gu16}  In 1975,  Heller derived the equations of motion for the four parameters (position ${\bf x}_0$, momentum ${\bf p}_0$, complex width matrix ${\hat \alpha}_0$, complex phase $\gamma_0$) of the GWP, assuming the PES is locally quadratic around ${\bf x}_0$, the \emph{Thawed Gaussian} approximation (TGA). The key of this method is that the phase-space center of the wavepacket moves by classical mechanics. That classical point is ``dressed" in the semiclassical width and phase. \cite{Heller75,Supp1}

In the adiabatic limit, the dynamics can be formally described in the framework of quantum mechanical description of the nuclei, ${\bf \Psi}({\bf x},t) = e^{i{H}({\bf x})t} {\bf \Psi}({\bf x},0)$ (here and in the following we set $\hbar=1$ unless stated otherwise), 
\begin{align}
\label{SE0}
{ H}({\bf x}) = -\sum_{i}^{N}{1\over 2m_i} {\partial^2\over \partial x_i^2} + {V}({\bf x}),
\end{align}
where {H}({\bf x})  is the Hamiltonian of the system. The potential {V}({\bf x}) is a parametric function of geometry $\bf x$, $m$ is the nuclear mass, and ${\bf \Psi}({\bf x},t)$ is the nuclear wavefunction. TGA can be alternatively derived by splitting the evolution operator operator $e^{-i{H}t}$ into slices $e^{-i{H}\epsilon}\,e^{-i{H}\epsilon}\,...$ with an infinitesimally small time step $\epsilon$.\cite{Supp1} If for a single time slice one expands $e^{i{H}({\bf x})\varepsilon}$ to first order in $\varepsilon$, applies the same approximation as Heller, and re-exponentiates, one recovers a new Gaussian with parameters shifted by one time step using Heller's equations of motion.\cite{Supp1,Heller75}  We seek to follow similar steps to generalize TGA for multiple electronic states.
\begin{figure}[t]
\vspace{0 cm}	
\includegraphics[width=9 cm]{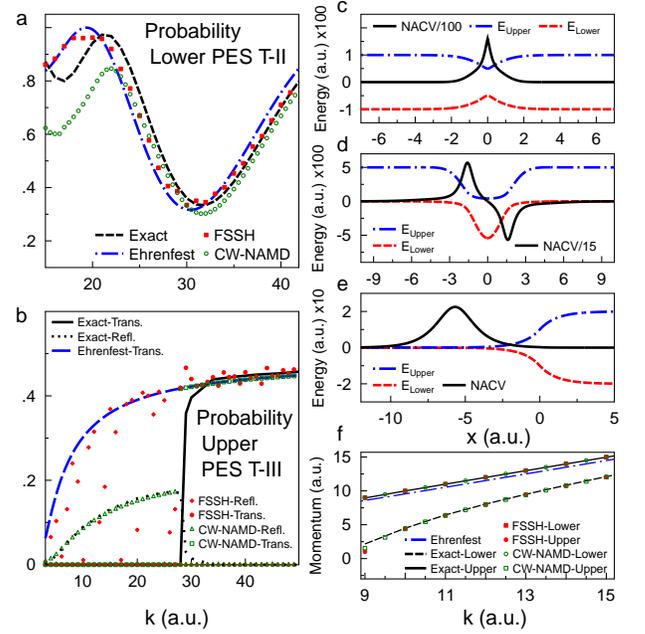}
\caption{\label{Tully} a (b)-  Scattering probabilities Tully II (III) problems on the Lower (Upper) surface for different initial wave vectors $k$. 
Exact solution, FSSH (2000 trajectories), Ehrenfest and  CW-NAMD are compared. Initial wavepacket position $x_{initial}=-10~a.u.$. Initial width, $\alpha_{initial}=ik^2/400$ for all. c (d,e)- Potential Energy Surfaces (E) and Non-Adiabatic Coupling Vectors (NACV) for Tully I (II, III).  g- Average momentum for each surface after scattering (Tully I). Exact solution, FSSH, Ehrenfest and RG-WP are compared. } 
\end{figure}

The non-adiabatic dynamics can similarly be obtained from a quantum mechanical description of the nuclei, $\vert {\bf \Psi}({\bf x},t) \rangle = e^{i{\hat H}({\bf x})t} \vert {\bf \Psi}({\bf x},0) \rangle$. Now the nuclei's potential energy operator $\hat{V}({\bf x})$ and the wavefunction $\vert \bf \Psi({\bf x},t) \rangle$ are $M\times M$ hermitian matrix and $M$ component vector respectively, where $M$ is the number of relevant electronic states. For simplicity we will consider a situation with two levels crossing, i.e., with $M=2$. This is the most common situation, typically more complex problems with multiple PESs and crossings can be modeled as consecutive transitions through well separated regions of coupling between two locally adjacent PESs. Furthermore, an extension to the $M>2$ situation is straightforward. We assume that the initial state is a single Gaussian localized on the first PES,       $|\Psi({\bf x},0)\rangle = N_0^{(1)} g^{(1)}({\bf x}; {\bf x}_0^{(1)}, {\bf p}_0^{(1)}, {\hat \alpha}_0^{(1)})|1[{\bf x}_0^{(1)}]\rangle$,
where $|1[{\bf x}_0^{(1)}]\rangle$ an eigenstate of ${\hat V}({\bf x})$ corresponding to the first PES evaluated at ${\bf x}_0^{(1)}$ and $N_0^{(1)}$ is the real amplitude of the otherwise normalized state $\vert {\bf \Psi}({\bf x},0)\rangle$. Here and in the following the superscripts indicate the electronic state or PES.  Non-Gaussian states can be treated as linear superpositions of finite number of Gaussians due to the linearity of the TDSE. 

We again split the evolution operator operator $e^{-i\hat{H}t}$ into slices $e^{-i\hat{H}\epsilon}\,e^{-i\hat{H}\epsilon}\,...$ and now introduce a basis resolution $\sum_{i=1,2} |i[{\bf x}_1^{(1)}]\rangle\langle i[{\bf x}_1^{(1)}]|$ between the first and the second slices (the subscripts here and below indicate the time steps). The point ${\bf x}_1^{(1)}$ is the location of the classical trajectory to be specified below. The use of the local electronic basis function is the first deviation from previously derived path-integral GWP dynamics.\cite{Krempl94,Coalson96} Physically introduction of the basis resolution corresponds to projecting the wavepacket on the new, slightly shifted basis of the eigenstates of $\hat V$ at the new average position of the wavepacket at time $\epsilon$. The new wavepacket will mostly remain in the electronic state $|1[{\bf x}_1^{(1)}]\rangle$ with a small ($\propto \epsilon$) transfer to $|2[{\bf x}_1^{(1)}]\rangle$. After some calculation one gets \cite{Supp1}

\begin{align}
\label{branch}
|\Psi({\bf x},\epsilon)\rangle = N_1^{(1)} g^{(1)} ({\bf x})|1[{\bf x}_1^{(1)}]\rangle + \epsilon \, N_1^{(2)}  g^{(2)} ({\bf x})|2[{\bf x}_1^{(1)}]\rangle.
\end{align}
The change in the wavepacket $g^{(1)}$({\bf x}) in Eq. \ref{branch} (i.e., after a single time step) is infinitesimal with the same form as the Heller GWP dynamics, leading to equations of motion for the multistate case:
\begin{align}
\label{EOMM}
&\dot{{\bf x}}_0^{(1)}={\bf p}^{(1)}_0\hat{m}^{-1}
\\ \nonumber
&\dot{{\bf p}}^{(1)}_0=-\langle 1[{\bf x}_0^{(1)}] \vert  {\partial}_{\bf x} {\hat V}({\bf x}_0) \vert 1[{\bf x}_0^{(1)}] \rangle,
\\ \nonumber
&\dot{{\hat \alpha}}^{(1)}_0=-2{\hat \alpha}^{(1)}_0\hat{m}^{-1}{\hat \alpha}^{(1)}_0 -\frac{1}{2}\langle 1[{\bf x}_0^{(1)}] \vert  {\partial}^2_{\bf x} {\hat V}({\bf x}_0) \vert 1[{\bf x}_0^{(1)}] \rangle,
\\ \nonumber
&\dot{\gamma}_0=i\hbar ~\text{Tr}[{\hat \alpha}_0\hat{m}^{-1}] + \frac{1}{2} {\bf p}_0\hat{m}^{-1}{\bf p}_0 -  \langle 1[{\bf x}_0^{(1)}] \vert {\hat V}({\bf x}_0) \vert 1[{\bf x}_0^{(1)}] \rangle .
\end{align} 
The weight, $ N_1^{(1)}=  N_0^{(1)} $, is unchanged. 

The wavepacket $g^{(2)}({\bf x})$ ``hopped" to PES 2. It has the same classical position as the original wavepacket, ${\bf x}_1^{(2)}={\bf x}_0^{(1)}$, but different momentum: ${\bf p}_1^{(2)}$ is such that ${\bf p}_1^{(2)} - {\bf p}_0^{(1)}$ is parallel to the non-adiabatic coupling vector ${\bf d}_{12}({\bf x}_0^{(1)})=\langle 2 [{\bf x}_0^{(1)}] \vert\partial_{\bf x} \vert  1 [{\bf x}_0^{(1)}] \rangle$, and its absolute value satisfies the energy conservation condition, $\sum_{\alpha=1}^N [(p_{1\alpha}^{(2)})^2 - (p_{0\alpha}^{(1)})^2]/(2m_\alpha)= E^{(1)}({\bf x}_0^{(1)})-E^{(2)}({\bf x}_0^{(1)}) $ \cite{Herman84-2,Supp1}. This rescaled momentum is a direct consequence of the projection onto local electronic basis functions. The parameters $\alpha^{(2)}$ and $N_1^{(2)}$  are related to the coefficients of $g^{(1)}({\bf x})$ as
\begin{align}
\label{alpha2}
&{\bf \alpha}_1^{(2)}  = {\bf \alpha}_0^{(1)} +\frac{1}{2}\frac { \langle 2[{\bf x}_0^{(1)}] \vert {\partial}^2_{\bf x}{\hat V}({\bf x}_0) \vert 1[{\bf x}_0^{(1)}] \rangle}{[{\bf d}^{(12)}({\bf x}^{(1)}_0)\cdot\bar{{\bf v}}^{(1)}_{1}]} \, ,
\\
\nonumber
&N_1^{(12)}  = N_0^{(1)}{\bf d}^{(12)}({\bf x}^{(1)}_0)\cdot\bar{{\bf v}}^{(1)}_{1}\,
\exp{\big[{{\bf d}^{(12)}({\bf x}^{(1)}_0)\cdot\Delta{\bf v}^{(1)}_{0}\over {\bf d}^{(12)}({\bf x}^{(1)}_0)\cdot \bar{{\bf v}}^{(1)}_{1}} \big]} ,
\end{align}
where $v_{0\alpha}^{(1)}=p_{0\alpha}^{(1)} /m_\alpha$, $\bar{{\bf v}}^{(1)}_{1}=({\bf v}^{(1)}_{0}+{\bf v}^{(2)}_{1})/2$ and $\Delta{\bf v}^{(1)}_{0}=({\bf v}^{(1)}_{0}-{\bf v}^{(2)}_{1})/2$. Note that the parameters of the spawned wavepacket, e.g. Eqs. \ref{alpha2}, change discontinuously at the moment of the hop.  In practice we only keep the linear term in the expansion of $\hat{V}({\bf x})$, since for realistic systems calculation of the quadratic term can be very costly. 

At the next time step each of the wavepackets propagates and spawns again, according to Eqs. \ref{branch} - \ref{alpha2} (with a replacement $1\rightarrow 2$ for the wavepacket on the second PES). After each time step the total number of the wavepackets doubles. Such process can be viewed as branching on a tree, shown in Fig. \ref{drawing}-a. This branching tree can be evaluated by a Monte-Carlo approach \cite{Gorshkov2013,White14,White15} which becomes too computationally expensive in systems with multiple level crossings. 

Here we propose a new approach based on the wavepacket reconstruction after each spawning event. The approach is schematically shown in Fig. \ref{drawing}-b. That is, after two time steps, described in Eqs. (\ref{branch} - \ref{alpha2}), one creates two wavepackets, on each PES, which will give rise to four more, etc. We note, however, that if each pair of the wavepackets on the same surface has close coordinates and momenta, one can replace each pair by a single GWP, with slightly shifted parameters. We parameterize the new Gaussian by calculating the expectation values of $\hat{x},\hat{p},\hat{x}^2,\hat{p}^2$ of the superposition. $\langle \hat{x} \rangle$ and $\langle \hat{p} \rangle$ are taken as the position and momentum of the RG wavepacket, while $\langle \hat{x}^2 \rangle$ and $\langle \hat{p}^2 \rangle$ directly give the new complex width. The new phase and weight, $\gamma$ and $N$, are determined by maximizing the overlap of the new wavepacket with the superposition, under the constraint that $\vert N \vert^2$ is the same as the density of the superposition.\cite{Supp1} Approximations are made in order to decouple the calculation of $\langle \hat{x} \rangle$ and $\langle \hat{p} \rangle$ from the explicit form of the wavefunction, $i.e. ~ \hat \alpha$.\cite{Supp1} Thus, as with Heller's equations, the trajectories of the GWPs remains independent of the phase and width. At the next step the procedure is repeated, again we have only two GWP and so on. The process repeats until the overlap, $O_{12}$, between the gaussians within each pair becomes intolerable, $O_{12}<O_{min}$. At this point, or if the non-adiabatic coupling drops below its own threshold, the ``coupling'' between the GWPs stops and each is treated independently, thus new branching is allowed. This coarse branching is schematically shown in Fig. \ref{drawing}-c. This approximation significantly reduces or eliminates the exponential growth of the number of wavepackets. We call this approximation coupled wavepackets for non-adiabatic molecular dynamics (CW-NAMD).

As the two wave packets separate in position space, their electronic bases will become non-orthogonal. Formally this must be taken into account by considering the required basis rotations when reconstruction occurs. These rotations lead to a correction, but it is small and does not affect the results presented in this letter.\cite{Supp1}

The CW-NAMD method is similar in spirit to the \emph{ab initio} multiple spawning (AIMS) method developed by Martinez \emph{et. al.}\cite{Martinez96,Nun00,Martinez06} Both involve approximate solution to an infinitely branching tree, of GWPs. However, in practice AIMS is usually based on independent, frozen GWP, which are non-interfering. CW-NAMD uses thawed GWPs and considers the full superposition of GWPs. For AIMS the ``spawning" procedure  is based on well-reasoned but empirical considerations,\cite{Levine08} and is only truly \emph{ab initio} in the limit of infinite spawning, so called Full AIMS. The branching procedure in CW-NAMD has a simple numerical control parameter, $O_{min}$. 
 
\begin{figure}[t]
\vspace{-0.5 cm}	
\includegraphics[width=9 cm]{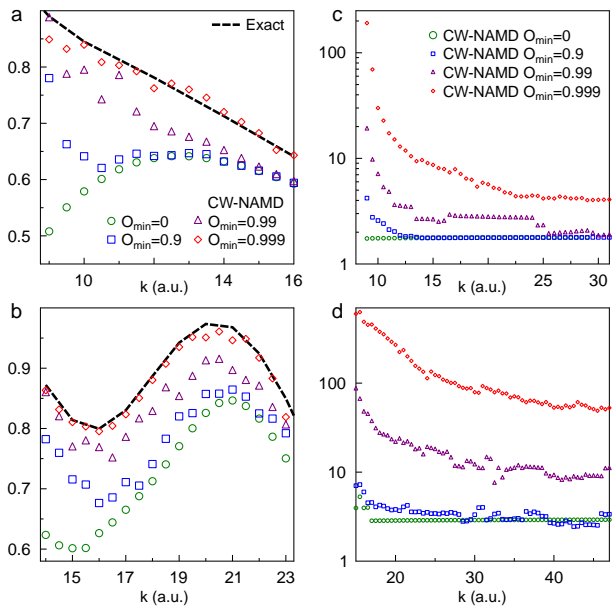}
\caption{\label{MX} a (b)- Comparison of low momentum transmission probabilities, on lower surface, with different values of $O_{min}$, compared to exact solution, for Tully I (II). Initial conditions as in Fig 2 (except $x_{initial}=-5~a.u.$ for Tully I). (c,d)- Number of ``effective" trajectories for Tully I (II) calculation with different values of $O_{min}$. Dynamics are run for a total time of $25,000 \over k$ ($40,000 \over k$ ) a.u. for Tully I (II).  } 
\end{figure}

Figure \ref{Tully} shows scattering probabilities for the standard Tully test problems II and III, Figure \ref{Tully}-d, e.   These problems are frequently used to test new methods of non-adiabatic dynamics because they specifically probe the interference (Tully II) and decoherence (Tully III) questions directly. We compare the CW-NAMD results, with $O_{min}=0$, to the standard fewest switching surface hopping (FSSH) and the mean-field Ehrenfest method as well as direct calculation of the time-dependent Schr{\" o}dinger equation. When branching does not occur, the computational cost of the CW-NAMD method is similar to Ehrenfest (i.e. there is one force calculation per surface per time point), and is much lower than surface hopping. Figure \ref{Tully} a,b demonstrates that for sufficiently high momentum the CW-NAMD method produces the correct scattering results. The CW-NAMD does not suffer from the interference or decoherence errors of Eherenfest or FSSH. This can be observed by comparing the position of the peaks of the Stueckelberg oscillations \cite{Stueckelberg} in Figure \ref{Tully}-a and the lack of false oscillations in the reflected probabilities in Figure \ref{Tully}-b.  Unlike Ehrenfest, CW-NAMD produces the correct momenta and positions of the wave packets on the upper and lower surface (see Figure \ref{Tully}-f). However at low momenta, the total scattering probability is not conserved and may be poorly estimated. This is evident in both Tully-II (see Figure \ref{Tully}-a, \ref{MX}-b) and Tully-I (\ref{MX}-a). This can be corrected by allowing the coupled GWPs to branch, \emph{i.e.} set $O_{min} > 0$. 

We compare the low momentum results for Tully I (II) with different values of $O_{min}$ in Figure \ref{MX} a (b). The difference between exact and CW-NAMD solutions is systematically improved by increasing  $O_{min}$. The increased cost can be seen in Figure \ref{MX} c (d). In direct dynamics simulations the bottleneck is typically the calculation of the PES gradients (forces). Trajectory methods like FSSH,  require one force calculation per time step per trajectory. Thus we define an ``effective" number of trajectories,  by determining the total number of force calculations (summed over all branches) divided by the total number of time steps for the simulation, to compare the cost of a branching scheme to that of a trajectory based methods (\emph{i.e.} FSSH). We see a growth of the number of trajectories required with increased $O_{min}$, However the cost of CW-NAMD is still lower than the 2000 trajectories used to calculate the FSSH result (Fig. \ref{Tully} b).  In the limit $O_{min}=1$ we recover the full branching tree (Fig. \ref{drawing} a). Lower values of $O_{min}$ result in a coarse-grained tree (Fig. \ref{drawing} c). To prevent overgrowth of the tree, we place hard-limits on the spawning rate and utilize pruning procedures to discard irrelevant branches. \cite{Supp1}

	In conclusion, the new CW-NAMD method is a highly efficient and accurate method of simulating non-adiabatic dynamics applicable to realistic molecular systems. CW-NAMD consistently accounts for decoherence and interference between different dynamical pathways. It can be as efficient as the Ehrenfest method in the high momentum limit, moreover it accurately describes the dynamics of branching wave packets. In the low momentum limit the method can be systematically improved by increase the rate of allowed branching via the user controlled accuracy threshold, $O_{min}$. Combined with filtering of insignificant branches, the method is more accurate and more efficient than the standard FSSH. In our test problems we observe numerical cost of CW-NAMD ranging from  about $2 (M)$ to $500$ trajectories depending on initial momentum and desired accuracy. This needs to be compared with the number of effective trajectories in other methods: $2(M)$ (Ehrenfest), $(2-5)\times 10^{3}$ (FSSH),  $(2-10)\times 10^{4}$ (Monte-Carlo approaches). The development of CW-NAMD opens new avenues for future research: more advance branching criterion, manipulation of the electronic bases, optimization of the reconstruction and branch pruning procedures, and application to molecular systems of increasing size. 
	
\begin{acknowledgments}
We acknowledge support of the U.S. Department of Energy through the Los Alamos National Laboratory (LANL) LDRD Program. LANL is operated by Los Alamos National Security, LLC, for the National Nuclear Security Administration of the U.S. Department of Energy under Contract No. DE-AC52- 06NA25396. We also acknowledge the LANL Institutional Computing (IC) Program provided computational resources. This work was supported in part by the Center for Nonlinear Studies (CNLS) and the Center for Integrated Nanotechnology (CINT) at LANL.
 \end{acknowledgments}

\bibliographystyle{apsrev4-1}
\end{document}


\begin{center}
{\bf \LARGE Supplemental Materials: Coupled Wavepackets for Non-Adiabatic Molecular Dynamics: A Generalization of Gaussian Wavepacket Dynamics to Multiple Potential Energy Surfaces}

\normalsize Alexander White, Sergei Tretiak, and Dmitry Mozyrsky

\normalsize \emph{Theoretical Division, Los Alamos National Laboratory, Los Alamos, New Mexico 87545, USA}

\end{center}

\section{Heller's Adiabatic Gaussian Wavepacket Dynamics from Path Integral approach}

Here we re-derive the \emph{Thawed Gaussian} approximation, originally derived by Heller, \cite{Heller75} based on a time slicing procedure. This will provide the foundation from which we will extend the derivation to multiple coupled potential energy surfaces. The wavefunction of the system at time $t$ is given by the time dependent Schr{\"o}dinger Equation and the initial wavefunction:
\begin{align}
\label{SE1}
\Psi({\bf x},t) =e^{-i\varepsilon {H}({\bf x})}e^{-i\varepsilon {H}({\bf x})} ... e^{-i\varepsilon {H}({\bf x})}e^{-i\varepsilon {H}({\bf x})}  \Psi({\bf x}, 0) ~.
\end{align}
Here we have defined $\varepsilon \equiv \frac{t}{M\hbar}$, where $M$ is a large number. For a single time step, we expand to first order in $\varepsilon$:

\begin{align}
\label{SE2}
\Psi({\bf x},t) =...  \big\{ 1 -i\varepsilon {K}({\bf x}) -i\varepsilon {V}({\bf x})  \big\}   \Psi({\bf x},0)  =
 \\
 \nonumber
...   \big\{ 1 - i\varepsilon \big( \sum_i \frac{-\hbar^2}{2m_i}\frac{\partial^2}{\partial{x_i}^2} + V({\bf x}) \big)\big\}\Psi({\bf x},0) 
\\ 
\nonumber
\approx ...  \Psi({\bf x},\hbar \varepsilon) ~.
\end{align}
Using similar notation to Heller the initial wavepacket is given by:\cite{Heller75}
\begin{align}
\label{gauss}
 \Psi({\bf x},0) =  \text{exp}[\,\frac{i}{\hbar}  \{\,\gamma_0 + {\bf p}_0^T({\bf x}-{\bf x}_0) + ({\bf x}-{\bf x}_0)^T{ {\hat \alpha}}_0({\bf x}-{\bf x}_0)\big\}] ~.
 \end{align}
For Equation \ref{SE2} we have a second derivative term:
\begin{align}
\label{KE}
&\sum_i \frac{-\hbar^2}{2m_i}\frac{\partial^2}{\partial{x_i}^2}  \Psi({\bf x},0) = \sum_i \frac{-i\hbar}{2m_i} \frac{\partial}{\partial{x}_i} \big\{ [{\bf p}_0]_i + [\,{\hat \alpha}_0({\bf x}-{\bf x}_0)]_i + [({\bf x}-{\bf x}_0)^T{\hat \alpha}_0]_i \big\}  \Psi({\bf x},0)
  \\
  \nonumber
& = \big\{ -i\hbar \, \text{Tr}[{\hat \alpha}_0 {\hat m}^{-1}] +   \frac{1}{2}{\bf p}_0^T {\hat m}^{-1} {\bf p}_0 + {\bf p}_0^T {\hat m}^{-1} {\hat \alpha}_0 ({\bf x}-{\bf x}_0) +
\\
\nonumber
&~~~~~~~~~~~~~~~~~~~~~~~~~~~~~~ ({\bf x}-{\bf x}_0)^T {\hat \alpha}_0  {{\hat m}}^{-1} {\bf p}_0  + 2\,({\bf x}-{\bf x}_0)^T{\hat \alpha}_0 {\hat m}^{-1} {\hat \alpha}_0 ({\bf x}-{\bf x}_0) \big\}  \Psi({\bf x},0) ~,
\end{align}
and terms from the potential expanded to the quadratic term around ${\bf x}_0$ (as in Ref.  \citenum{Heller75}):
\begin{align}
\label{V}
V({\bf x})\approx V({\bf x}_0) + {\bf V'}({\bf x}_0)^T ({\bf x}-{\bf x}_0) + \frac{1}{2}({\bf x}-{\bf x}_0)^T {\hat V''}({\bf x}_0)  ({\bf x}-{\bf x}_0) ~.
\end{align}
All terms in Eqs. \ref{KE} and  \ref{V} are of order $\varepsilon \hbar$. Thus they can be collected and returned to the exponential form (accurate up to order $\varepsilon$):
\begin{align}
\label{newPsi}
 \Psi({\bf x},\varepsilon\hbar) \approx \text{exp} \Big[ \frac{i}{\hbar} &\Big\{ \gamma_0 + \varepsilon \hbar ( i\hbar \, \text{Tr}[{\hat \alpha}_0 {\hat m}^{-1}]   +   \frac{1}{2}{\bf p}_0 {\hat m}^{-1} {\bf p}_0 - V({\bf x}_0)   )
\\
\nonumber
&+ \big\{ {\bf p}_0 -\varepsilon\hbar  {\bf V'}({\bf x}_0) \big\}^T({\bf x}-{\bf x}_0 - \varepsilon\hbar{\hat m}^{-1} {\bf p}_0 )
\\
\nonumber
+ ({\bf x}-{\bf x}_0 - \varepsilon\hbar {\hat m}^{-1} {\bf p}_0 )^T&\big\{{\hat \alpha}_0-\varepsilon\hbar \big( 2 {\hat \alpha}_0 {\hat m}^{-1} {\hat \alpha}_0 + \frac{ {\hat V''}}{2} \big)
 \big\}({\bf x}-{\bf x}_0 - \varepsilon\hbar{\hat m}^{-1} {\bf p}_0 ) \Big\} + O(\varepsilon^2)\Big] ~.
\end{align}
By defining updated Gaussian variables we now have equations of motion which are accurate up to first order in the time step $dt=\varepsilon\hbar$:
\begin{align}
\label{x}
\dot{\bf x}_0&={\hat m}^{-1} {\bf p}_0 ~,
\\
\label{p}
\dot{\bf p}_0&= -{\bf V'}({\bf x}_0) ~,
\\
\dot{\hat \alpha}_0&= -2 {\hat \alpha}_0 {\hat m}^{-1} {\hat \alpha}_0 - \frac{ {\hat V''}}{2} ~,
\\
\dot\gamma_0&=i\hbar \, \text{Tr}[{\hat \alpha}_0 {\hat m}^{-1}]   +   \frac{1}{2}{\bf p}_0 {\hat m}^{-1} {\bf p}_0 - V({\bf x}_0) ~.
\end{align}
This result is exactly that of Heller's multidimensional thawed Gaussian wavepacket dynamics.\cite{Heller75}
\section{Generalization to multi-state system}
We again begin with the time dependent Schr{\"o}dinger Equation. This time our wavefuncton is a vector, and the Hamiltonain is a matrix, in electronic state space (defined for a specific geometry $\bf x$):
\begin{align}
\label{SE3}
\vert \Psi({\bf x},t) \rangle =e^{-i\varepsilon \hat{H}({\bf x})}e^{-i\varepsilon \hat{H}({\bf x})} ... e^{-i\varepsilon \hat{H}({\bf x})}e^{-i\varepsilon \hat{H}({\bf x})} \vert \Psi({\bf x}, 0) \rangle ~.
\end{align}
The initial wavefunction can undergo a \emph{Born-Oppenheimer expansion}, where for each ${\bf x}$ the wavefunction is expanded in a basis of eigenstates of $\hat{V}({\bf x})$:
\begin{align}
\label{BOE}
\vert \Psi({\bf x}, 0) \rangle= \sum_n \vert n[{\bf x}] \rangle \langle n[{\bf x}]   \vert \Psi({\bf x},0)  \rangle ~.
\end{align} 
However, here we will take a different approach. For all {\bf x} we will expand in the eigenstates of $\hat{V}({\bf x})$ at ${\bf x}= {\bf x}_0$, where ${\bf x}_0$ is the center of the Gaussian wavepacket $\langle n[{\bf x}]  \vert \Psi({\bf x},0)  \rangle$:
\begin{align}
\label{BOE2}
\vert \Psi({\bf x}, 0) \rangle= \sum_n \vert n[{\bf x}_0] \rangle \langle n[{\bf x}_0]   \vert \Psi({\bf x},0)  \rangle ~.
\end{align} 
 While this choice if formally legal, $\sum_n \vert n[{\bf x}_0] \rangle \langle n[{\bf x}_0]   \vert = 1$, it may seem a strange choice. However, it is fully consistent with the trajectory based branching scheme that will ultimately be used to solve this system of equations, and gives the correct receipt for ``hopping" trajectories' boundary conditions. We will consider the wavepacket at time $\hbar\varepsilon$ on state $m$, which is projected onto the basis of eigenstates of $\hat{V}({\bf x}_1)$, where ${\bf x_1}$ is the center of the wavepacket at time $\hbar\varepsilon$. Taking only the first $\varepsilon$ step in Eq. \ref{SE3}, and expanding the exponential to first order in $\varepsilon$ we have:
 \begin{align}
 \label{SE4b}
  \langle m [{\bf x}_1] \vert \Psi({\bf x},\hbar \varepsilon) \rangle \approx  \sum_n \langle m  [{\bf x}_1] \vert \big\{ \hat{I} -i\varepsilon \hat{K}({\bf x}) -i\varepsilon \hat{V}({\bf x})  \big\} \vert n[{\bf x}_0] \rangle \langle n[{\bf x}_0]   \vert \Psi({\bf x},0)  \rangle ~.
\end{align}
 Eq. \ref{SE4b} describes the wavepacket at time $\hbar\varepsilon$ on electronic surface $m$, which has contributions from wavepackets at time $0$ on all surfaces ($n$). The eigenstate $\vert m [{\bf x}_1] \rangle$ can be projected in the basis of $\vert \chi [{\bf x}_0] \rangle$ eigenstates (up to the first order in $\varepsilon$):
\begin{align}
\label{zetaexp}
\vert m \rangle= \sum_l  \vert l \rangle\langle l \vert m \rangle = \sum_l \big\{ \delta_{l,m}   + \langle l \vert \nabla_{{\bf x}_1} m\rangle \cdot ({\bf x}_1 - {\bf x}_0)  \big\}\vert l \rangle
+... 
\\
\nonumber
\approx  \vert m \rangle + \varepsilon\hbar \sum_l  {\bf d}_{l,m} \cdot {\hat m}^{-1} {\bf p}_0 \vert l \rangle +O(\varepsilon^2) ~.
\end{align}
We have assumed that the difference ${\bf x_1}-{\bf x_0}$ is proportional to $\hbar\varepsilon$ and that we have constant momentum over that time step. ${\bf p}_0$ is the momentum of $\langle n[{\bf x}_0]   \vert \Psi({\bf x},0)  \rangle$.
 From this point we will drop the $[{\bf x}_0]$ label and assume, unless otherwise labeled, that our electronic basis states are eigenstates of $\hat{V}({\bf x}_0)$. 

If we insert Eq. \ref{zetaexp} into Eq. \ref{SE4b}:
\begin{align}
  \langle m [{\bf x}_1] \vert \Psi({\bf x},\hbar \varepsilon) \rangle \approx  \sum_{n}  \Big[ \langle m \vert  & +\hbar\varepsilon\sum_l {\bf d}_{l,m} \cdot  {\hat m}^{-1} {\bf p}_0 \langle l \vert  \Big]\times
\\
\nonumber  
&\big\{ 1 - i\varepsilon \big( \hat{K}({\bf x}) + \hat{V}({\bf x})\big)\big\}  \vert  n \rangle \langle n \vert \Psi({\bf x},0) \rangle ~,
\end{align}
and separate out the term $m=n$ in the sum, keep only terms up to $O(\varepsilon$), we have:
\begin{align}
\label{sep}
& \langle m [{\bf x}_1] \vert \Psi({\bf x},\hbar \varepsilon) \rangle \approx  
\\
\nonumber
&\langle m \vert \big\{ \hat{I} -i\varepsilon \hat{K}({\bf x}) -i\varepsilon \hat{V}({\bf x})  \big\} \vert m\rangle \langle m  \vert \Psi({\bf x},0)  \rangle 
\\
\nonumber
& -i\varepsilon   \sum_{n\neq m} \Big[ i\hbar{\bf d}_{n,m} \cdot  {\hat m}^{-1} {\bf p}_0
+\langle m \vert \big\{\hat{K}({\bf x}) + \hat{V}({\bf x})  \big\} \vert n\rangle \Big]\langle n  \vert \Psi({\bf x},0)  \rangle ~.
\end{align}
Here we have used the fact that ${\bf d}_{m,m}=0$ and $\vert m/n \rangle$ is an eigenstate in order to reduce terms. We now expand the potential energy matrix operator around ${\bf x}_0$:
\begin{align}
\label{Vop}
\hat{V}({\bf x})&\approx \hat{V}({\bf x}_0) + \hat{\bf V}'({\bf x}_0)^T ({\bf x}-{\bf x}_0) + \frac{1}{2}({\bf x}-{\bf x}_0)^T \hat{\bf \underline V}''({\bf x}_0)  ({\bf x}-{\bf x}_0)
\\
\nonumber
&=\sum_{\alpha,\beta} \Big\{ {V}_{\alpha,\beta}({\bf x}_0) + {\bf V}'_{\alpha,\beta}({\bf x}_0)^T ({\bf x}-{\bf x}_0) + \frac{1}{2}({\bf x}-{\bf x}_0)^T {\bf \underline V}''_{\alpha,\beta}({\bf x}_0)  ({\bf x}-{\bf x}_0)\Big\} \vert \alpha \rangle \langle \beta \vert ~,
\end{align}
where the $\alpha/\beta$ electronic basis set is ${\bf x_0}$ independent, $e.g$ the atomic orbital basis set in realistic calculations, or the diabatic basis set in most model problems. Inserting Eq. \ref{Vop} into \ref{sep} leads to:
\begin{align}
\label{sep-exp}
& \langle m [{\bf x}_1] \vert \Psi({\bf x},\hbar \varepsilon) \rangle \approx  
\\
\nonumber
& \big\{ 1 -i\varepsilon \big[ \sum_i \frac{-\hbar^2}{2M_i}\frac{\partial^2}{\partial{x_i}^2}  + E_m({\bf x}_0) - {\bf F}_m({\bf x}_0)^T ({\bf x}-{\bf x}_0)+ ({\bf x}-{\bf x}_0)^T\tilde{\bf \underline H}_m({\bf x}_0)({\bf x}-{\bf x}_0)\big]  \big\}  \langle m  \vert \Psi({\bf x},0)  \rangle 
\\
\nonumber
& -i\varepsilon   \sum_{n\neq m} \Big[ i\hbar{\bf d}_{n,m} \cdot {\hat m}^{-1} {\bf p}_0 + {\bf V}_{m,n}'({\bf x}_0)^T({\bf x}-{\bf x}_0)+ \frac{1}{2}({\bf x}-{\bf x}_0)^T{\bf \underline V}_{m,n}''({\bf x}_0)({\bf x}-{\bf x}_0)  \big\}  \Big]\langle n  \vert \Psi({\bf x},0)  \rangle ~.
\end{align}

Now we consider the form
\begin{align}
\langle n  \vert \Psi({\bf x},0)  \rangle   \equiv  \text{exp}[\,\frac{i}{\hbar}  \{\,\gamma_0 + {\bf p}_0^T({\bf x}-{\bf x}_0) + ({\bf x}-{\bf x}_0)^T{\hat \alpha}_0({\bf x}-{\bf x}_0)\big\}] \times N_n ~.
\end{align}

The first line ($m=n$ case) of Eq. \ref{sep-exp} is nearly identical to Eq. \ref{SE2} with some minor differences. First, the real weight of the wavepacket ($N_n$) arises since the initial state need not be pure. Second while the first expansion term is the force vector $\langle m \vert\hat{\bf V}'({\bf x}_0)\vert m \rangle= \frac{\partial}{\partial{{\bf x}}}E_m({\bf x})\vert_{{\bf x}={\bf x}_0}=-{\bf F}_m({\bf x}_0)$ due to the Hellman Feynman theorem, the second expansion term is not the true Hessian matrix of the potential energy surface ($\tilde{\bf \underline H}_m({\bf x}_0)   \neq \frac{\partial^2}{\partial{\bf x}^2}E_m({\bf x})\vert_{{\bf x}={\bf x}_0})$, because we first expanded in the basis which is ${\bf x_0}$ invariant, then rotated into the eigenbasis of $\hat{V}({\bf x}_0)$.  Thus we have:
\begin{align}
\label{Hess}
\sum_{\alpha,\beta} \langle m \vert \alpha \rangle  {\bf \underline V}''_{\alpha,\beta} ({\bf x}_0)\langle \beta \vert m \rangle \equiv \tilde{\bf \underline H}_{m}({\bf x}_0) ~.
\end{align}
Finally there is an additional zeroth order in (${\bf x}-{\bf x}_0$) term, coming from the $K_{m,m}({\bf x}_0)$.
\begin{align}
\label{K1}
K_{m,m}({\bf x}_0) \langle m  \vert \Psi({\bf x},0)  \rangle =- \frac{\hbar^2}{2m_i} \frac{\partial^2}{\partial{x}^2_i} \langle m  \vert \Psi({\bf x},0)\rangle ~.
\end{align}
Thus this first line tells us that the diagonal term is just Heller's \emph{Thawed Gaussian} wavepacket dynamics, but with the second derivative matrix given by Eq. \ref{Hess}. 

Now we turn our attention to the $m\neq n$ terms ${\bf V}_{m,n}'({\bf x}_0)$ and  $\hat{\bf \underline V}_{m,n}''({\bf x}_0)$. The first derivative term is related to the non-adiabatic coupling vectors through the Hellman-Feynman theorem:
\begin{align}
\label{NAC}
{\bf V}_{m,n}'({\bf x}_0)= \langle m [{\bf x}] \vert \hat{\bf V}'({\bf x} ) \vert   n [{\bf x}] \rangle \Big \vert_{{\bf x}={\bf x}_0}=  {\bf d}_{n,m}^T [E_m({\bf x}_0) - E_n({\bf x}_0)] ~.
\end{align}
The second derivative is similar to the diagonal case:
\begin{align}
\label{NASD}
{\bf \underline V}_{m,n}''({\bf x}_0) =\sum_{\alpha,\beta} \langle m \vert \alpha \rangle  {\bf \underline V}''_{\alpha,\beta} ({\bf x}_0)\langle \beta \vert n \rangle ~.
\end{align}

Collecting all the terms, up to $O(\varepsilon)$, and defining $D_{n,m}\equiv {\bf d}_{n,m}^T {\hat m}^{-1} {\bf p}_0$ we have:
\begin{align}
 \langle m [{\bf x}_1] \vert \Psi({\bf x},\hbar \varepsilon) \rangle \approx  =&... +   \hbar\varepsilon\sum_{n\neq m}  D_{n,m}\Big[ 1 + \frac{i}{\hbar} D^{-1}_{n,m}[E_n({\bf x}_0) - E_m({\bf x}_0)]  {\bf d}_{n,m}^T({\bf x} - {\bf x}_0) 
\\\nonumber
 &  - \frac{i}{\hbar}  \frac{1}{2} ({\bf x}-{\bf x}_0)^T D^{-1}_{n,m}{\hat V}_{m,n}''({\bf x}_0) ({\bf x}-{\bf x}_0)   \Big]  \langle n  \vert \Psi({\bf x},0)\rangle  ~.
\end{align}
 We can reduce the  expression to :
\begin{align}
 \langle m [{\bf x}_1] \vert \Psi({\bf x},\hbar \varepsilon) \rangle \approx  &\equiv ... + \hbar \varepsilon  \sum_{n\neq m}  D_{n,m}\Big[ 1 +\frac{i}{\hbar}{\bf \Delta P}_{m,n}^T({\bf x}-{\bf x}_0) +\frac{i}{\hbar} ({\bf x}-{\bf x}_0) ^T {\Delta \hat \alpha}_{m,n} ({\bf x}-{\bf x}_0)\Big] ~.
\end{align}
The term inside the bracket can be exponentiated (assuming all terms are small):
 \begin{align}
\text{exp}\Big\{ &\frac{i}{\hbar}{\bf \Delta P}_{m,n}^T({\bf x}-{\bf x}_0) +\frac{i}{\hbar}({\bf x}-{\bf x}_0) ^T{\Delta \hat \alpha}_{m,n} ({\bf x}-{\bf x}_0)\Big\}
\\
\nonumber
&\approx \Big[1+ \frac{i}{\hbar}{\bf \Delta P}_{m,n}^T({\bf x}-{\bf x}_0) + \frac{i}{\hbar}({\bf x}-{\bf x}_0) ^T{\Delta \hat \alpha}_{m,n} ({\bf x}-{\bf x}_0) \Big] ~,
 \\\nonumber
 &\text{where} ~{\Delta \hat \alpha}_{m,n} = -\frac{1}{2}\frac{{\hat V}_{m,n}''({\bf x}_0)}{{\bf d}_{n,m}^T {\hat m}^{-1}{\bf p}_0} ~,
 \\\nonumber
&\text{and}~~~~{\bf \Delta P}_{m,n} =  \frac{[E_n({\bf x}_0) - E_m({\bf x}_0)]}{{\bf d}_{n,m}^T {\hat m}^{-1}{\bf p}_0} {\bf d}_{n,m}^T ~.
\end{align}
This condition for the shift in momentum on hop is the same as previously derived. It conserves energy approximately (exactly in infinitely high momentum limit). To ensure that all trajectories conserve energy exactly for all momenta, we make the approximate transformation ${\bf d}_{n,m}^T {\hat m}^{-1} {\bf p}_0 = {\bf d}_{n,m}^T {\hat m}^{-1} \{{\bf p}_0+{\bf p}_1\}/2 + {\bf d}_{n,m}^T {\hat m}^{-1} \{{\bf p}_0-{\bf p}_1\} /2 \approx  {\bf d}_{n,m}^T {\hat m}^{-1} \{{\bf p}_0+{\bf p}_1\}/2 \times \text{exp}\Big[\frac{{\bf d}_{n,m}^T \{{\bf p}_0-{\bf p}_1\}}{{\bf d}_{n,m}^T \{{\bf p}_0+{\bf p}_1\} }\Big]$. With this consideration, and assuming ${\bf d}_{n,m}^T \{{\bf p}_0-{\bf p}_1\} << {\bf d}_{n,m}^T \{{\bf p}_0+{\bf p}_1\}$ our final result for the ($m \neq n$) case:
\begin{align}
\label{final}
 \langle m [{\bf x}_1] \vert \Psi({\bf x},\hbar \varepsilon) \rangle &\approx ... + \sum_{n\neq m}  \frac{1}{2} {\bf d}_{n,m}^T {\hat m}^{-1}\{ {\bf p}_0 + {\bf p}_1\} \times \text{exp}\Big[\frac{{\bf d}_{n,m}^T \{{\bf p}_0-{\bf p}_1\}}{{\bf d}_{n,m}^T \{{\bf p}_0+{\bf p}_1\} }\Big] \times N_n 
 \\\nonumber
 &~~~~~~~~~~~ \times \text{exp}[\,\frac{i}{\hbar}  \{\,\gamma_0 + {\bf p}_1^T({\bf x}-{\bf x}_0) + ({\bf x}-{\bf x}_0)^T{\hat \alpha}_1({\bf x}-{\bf x}_0)\big\}] ~,
 \\\nonumber
 &\text{where}~ {\bf p}_1= {\bf p}_0 + \frac{[E_n({\bf x}_0) - E_m({\bf x}_0)]}{{\bf d}_{n,m}^T {\hat m}^{-1}\{ \frac{{\bf p}_0 + {\bf p}_1}{2}\}} {\bf d}_{n,m}^T~,
 \\\nonumber
 &\text{and}~~~~ {\hat \alpha}_1= {\hat \alpha}_0 + \frac{{\hat V}_{m,n}''}{{\bf d}_{n,m}^T {\hat m}^{-1}\{ {\bf p}_0 + {\bf p}_1\}} ~.
\end{align}

\section{Wavepacket Reconstruction}
\label{WPR}
For Coupled Propagation and GWP Consolidation (See Section \ref{SOA}) we need to calculate overlap of two normalized Gaussians. We seek to define a single Gaussian which closely approximates two separate but similar Gaussians:
\begin{align}
N_Ge^{i\gamma_G}\vert G ({\bf x}_G, {\bf p}_G, {\hat \alpha}_G) \rangle \approx  N_1e^{i\gamma_1}\vert g_1 ({\bf x}_1, {\bf p}_1, {\hat \alpha}_1) \rangle  + N_2e^{i\gamma_2}\vert g_2 ({\bf x}_2, {\bf p}_2, {\hat \alpha}_2) \rangle \equiv \vert  \psi \rangle
\end{align} where $G,g_1$ and $g_2$ are normalized unphased complex Gaussians.
In the limit that the superposition $\vert \psi \rangle$ is indeed a Gaussian then the mapping is exact: 
\begin{align}
\label{NG}
&N_G=\vert {\langle   \psi \vert  \psi \rangle} \vert ^\frac{1}{2} ~,
\\
\label{Xg}
&{\bf x}_G= \frac{ \langle   \psi \vert {\bf x} \vert  \psi \rangle}{\langle   \psi \vert  \psi \rangle}\, ,
\\
\label{Pg}
&{\bf p}_G= -i \frac{ \langle   \psi \vert {\partial}_{\bf x} \vert  \psi \rangle}{\langle   \psi \vert  \psi \rangle}\, ,
\\
&\text{Im}[\hat \alpha_G] = \frac{1}{4} \Big [ \frac{ \langle   \psi \vert ({\bf x}-{\bf x}_G)^2 \vert  \psi \rangle}{\langle   \psi \vert  \psi \rangle} \Big ]^{-1}\, ,
\\
\label{re-alpha}
& \frac{ \langle   \psi \vert (-i{\partial}_{\bf x}-{\bf p}_G)^2 \vert  \psi \rangle}{\langle   \psi \vert  \psi \rangle} - \text{Im}[\hat \alpha_G] = \text{Re}[\hat \alpha_G] \, \text{Im}[\hat \alpha_G]^{-1} \text{Re}[\hat \alpha_G]\, .
\end{align}
Equation \ref{re-alpha} can be solved using the Geometric mean for positive definite matrices $\text{Im}[\hat \alpha_G]$ and $\frac{ \langle   \psi \vert (-i{\partial}_{\bf x}-{\bf p}_G)^2 \vert  \psi \rangle}{\langle   \psi \vert  \psi \rangle} - \text{Im}[\hat \alpha_G]$:
\begin{align}
\text{Re}[\hat \alpha_G]= \text{Im}[\hat \alpha_G]^{\frac{1}{2}}\Big[ \text{Im}[\hat \alpha_G]^{-\frac{1}{2}} \{\frac{ \langle   \psi \vert (-i{\partial}_{\bf x}-{\bf p}_G)^2 \vert  \psi \rangle}{\langle   \psi \vert  \psi \rangle} - \text{Im}[\hat \alpha_G] \} \text{Im}[\hat \alpha_G]^{-\frac{1}{2}} \Big]^\frac{1}{2} \text{Im}[\hat \alpha_G]^{\frac{1}{2}} \, .
\end{align}
These expectation values and overlap of a superposition of multivariate Gaussians can be calculated analytically. Finally the phase can be found by maximizing the overlap of $\langle \psi \vert N_Ge^{i\gamma_G}\vert G ({\bf x}_G, {\bf p}_G, {\hat \alpha}_G) \rangle$, under the constraint of Equation \ref{NG}:
\begin{align}
\label{SG}
e^{i\gamma_G}= \frac{\langle  G ({\bf x}_G, {\bf p}_G, {\hat \alpha}_G) \vert \psi \rangle  }{\vert \langle \psi \vert G ({\bf x}_G, {\bf p}_G, {\hat \alpha}_G) \rangle \vert } ~.
\end{align}
The expectation values of the superposition are given as a sum over combinations of the Gaussians:
\begin{align}
\langle \psi \vert O \vert\psi \rangle = \sum_{i,j \in 1,2} \langle g_j \vert O \vert g_i \rangle N_i N_j \text{exp}[i\{\gamma_i-\gamma_j\}] \, .
\end{align}       

Through Equations \ref{Xg}-\ref{Pg} the dynamics of the Coupled GWP depends strongly on the Thawed Gaussian Approximation, and the particular value of ${\hat \alpha}$. As the Coupled GWPs separate and the approximation of Equations \ref{NG}-\ref{SG} become less valid, the dependence of  Equations \ref{Xg}-\ref{Pg} on  ${\hat \alpha}$ can lead to unstable dynamics. This is particularly true when the value of $\text{Im}[{\hat \alpha}]$ becomes small (a very wide wavepacket). One valuable feature of the Thawed Gaussian Approximation is that the dynamics are fully classical, and independent of the phase and width. In the same spirit, here we seek to add further approximations to the dynamics which will add stability, with minimal sacrifice of accuracy. We note that typically we will break coupling when $\vert \langle g_1 \vert g_2 \rangle \vert$ is much less than unity.  The quantity $\langle g_1 \vert g_2 \rangle$ appears in the evaluation of all Equations \ref{NG}-\ref{SG}.  
\begin{align}
\langle g_1 \vert g_2 \rangle= (\text{Det}[{\frac{2}{\pi}{\hat \alpha}_2}])^{\frac{1}{4}} (\text{Det}[{\frac{2}{\pi}{\hat \alpha}_1}])^{\frac{1}{4}}  \int {\bf dx} \,  &\text{exp}\big[i\{ ({\bf x} - {\bf x}_2) {\hat \alpha}_2 ({\bf x} - {\bf x}_2) -  ({\bf x} - {\bf x}_1) {\hat \alpha}^*_1 ({\bf x} - {\bf x}_1)\}  \big ]
\\
\nonumber
&\times \text{exp}\big[i\{ {\bf p}_1 {\bf x}_1 - {\bf p}_2 {\bf x}_2\}   \big] \times \text{exp}\big[i\{ {\bf p}_2 - {\bf p}_1\} {\bf x}   \big].
\end{align} 
To stabilize the dynamics we make approximations when finding ${\bf x}_G$ and ${\bf p}_G$. For high momentum and close narrow wavepackets:
\begin{align}
\nonumber
(\text{Det}[{\frac{2}{\pi}{\hat \alpha}_2}])^{\frac{1}{4}} (\text{Det}[{\frac{2}{\pi}{\hat \alpha}_1}])^{\frac{1}{4}}   \text{exp}\big[i\{ ({\bf x} - {\bf x}_2) &{\hat \alpha}_2 ({\bf x} - {\bf x}_2) -  ({\bf x} - {\bf x}_1) {\hat \alpha}^*_1 ({\bf x} - {\bf x}_1)\}  \big ] \approx
\delta({\bf x}-[\frac{{\bf x}_1+{\bf x}_2}{2}]) ~,
\end{align}
which leads to:
\begin{align}
\label{apolap}
\langle g_1 \vert g_2 \rangle \approx \text{exp}\big[i\{\frac{{\bf p}_2 + {\bf p}_1}{2}\} \{ {\bf x}_1-{\bf x}_2\} \big] ~.
\end{align}

Use of Equation \ref{apolap} leads to:
\begin{align}
\label{XA}
\langle g_1 \vert {\bf x} \vert g_2 \rangle = -i \partial_{{\bf p_2}} \langle g_1\vert g_2 \rangle + \langle g_1 \vert {\bf x}_1 \vert g_2 \rangle &\approx   \frac{{\bf x}_1+{\bf x}_2}{2} \text{exp}\big[i\{\frac{{\bf p}_2 + {\bf p}_1}{2}\} \{ {\bf x}_1-{\bf x}_2\} \big] ~,
\\
\nonumber
-i\langle g_1 \vert { \partial_{\bf x}} \vert g_2 \rangle = -\frac{i}{2}\big[ \langle g_1 \vert { \partial_{\bf x}} g_2 \rangle - \langle { \partial_{\bf x}} g_1 \vert g_2 \rangle \big ] &= \frac{{\bf p}_1 +  {\bf p}_2}{2}  \langle g_1\vert g_2 \rangle + {\hat \alpha}_2 \langle g_1 \vert {\bf x} -{\bf x}_2 \vert g_2 \rangle + {\hat \alpha}_1^* \langle g_1 \vert {\bf x} -{\bf x}_1 \vert g_2 \rangle
\\
\label{PA}
&\approx \frac{{\bf p}_1 +  {\bf p}_2}{2}\text{exp}\big[i\{\frac{{\bf p}_2 + {\bf p}_1}{2}\} \{ {\bf x}_1-{\bf x}_2\} \big] ~.
\end{align}
Equations  \ref{XA} and \ref{PA} are used to in the calculations shown in the main text. In the calculation of Equation \ref{PA} we discard the terms which are proportional to ${\hat \alpha}$. This is consistent with the approximation leading to Equation \ref{apolap}, that momentum is high and wave packets are narrow $\big (\frac{{\bf p}_1 +  {\bf p}_2}{2} \gg \frac{1}{2} \{{\bf x}_1 -  {\bf x}_2\} \{{\hat \alpha}_2 - {\hat \alpha}_1^* \} \big)  $. For the calculation of ${\hat \alpha}_G$ the full set of Equations \ref{NG}-\ref{re-alpha} are used, but with the GWPs following the dynamics guided by Equations \ref{XA} and \ref{PA}. Using Equation \ref{apolap} in the evaluation of $N_G$ and $\gamma_G$ (Equations \ref{NG} and \ref{SG}) provides similar results to using the full $\langle g_1 \vert g_2 \rangle$ calculation for the models considered in the main text.


\section{Summary of the Algorithm}
\label{SOA}
\begin{description}
 \item [Initial Propagation] \hfill \\
\begin{itemize}
  \item[\emph{Step 1}:] Initialize Gaussian wavepacket (GWP) with desired parameters, on state $m$.
  \item[\emph{Step 2}:] Propogate the GWP forward in time using Eq. 5 from the main text. 
  \item[\emph{Step 3}:] If outside region of non-adiabatic coupling (NAC) repeat step 2. If GWP reaches a region of significant NAC, ${\bf d}_{n,m}^T \frac{{\hat m}^{-1} \{{\bf p}_0+{\bf p}_1\}/2}{\vert {\hat m}^{-1} \{{\bf p}_0+{\bf p}_1\}/2 \vert} > D_{min}$ (a user set threshold), then generate new wavepacket on state $n$ using Eq. 6 from main text and energy conserving change in momentum. If energy cannot be conserved, generation is not allowed. Establish a connection between these GWPs on $m$ and $n$, and begin Coupled Propogation.
\end{itemize}
 \item [Coupled Propogation] \hfill \\
\begin{itemize}
  \item[\emph{Step 4}:] Propogate all GWP forward in time using Eq. 5 from main text. 
  \item[\emph{Step 5}:] Calculate overlap of  normalized GWP with it's connections normalized ``hopped" GWP (with width and momentum shift described above). If the amplitude of the overlap is greater than an accuracy controlling threshold, $O_{min}$, reconstruct new wavepacket as described in Section \ref{WPR} . If the amplitude of the overlap is lower than $O_{min}$, eliminate connection between these GWPs. Continue with Step 2 for each wavepacket independently. 
  \item[\emph{Step 6}:]  If a GWP leaves the region of significant NAC, ${\bf d}_{n,m}^T \frac{{\hat m}^{-1} \{{\bf p}_0+{\bf p}_1\}/2}{\vert {\hat m}^{-1} \{{\bf p}_0+{\bf p}_1\}/2 \vert} < D_{min}$, eliminate connection between these GWPs. Continue with Step 2 for each wavepacket independently. 
\end{itemize}
 \item [End of Propogation] \hfill \\
\begin{itemize}
  \item[\emph{Step 7}:] Once final (or output) time is reached, the wavefunction is given as a sum over the weighted complex GWPs. Expectation values can be calculated directly from the wavefunction. 
\end{itemize}
 \item [Consolidation and Filtering (Optional Consideration)] \hfill \\
 Periodically, on some predefined number of time step, one could attempt to condense the trajectories by using the same reconstruction method as used in coupled propagation. Additionally after such a consolidation step one could discard trajectories which are insignificant (by real weight).  
 \begin{itemize}
   \item[\emph{Step A}:] Remove connection between all coupled trajectories. 
   \item[\emph{Step B}:] Loop through PES,
   \item[\emph{Step C}:] Find the largest weight GWP on the PES, this GWP is the initial value of the reconstructed GWP (GWP-New). 
   \item[\emph{Step D}:] Loop through all available GWPs on the PES. If the normalized GWP has overlap with the normalized GWP-New is higher than $O_{min}$, then add the GWP to GWP-New (by reconstruction). Continue to update GWP-New until fully looped through the GWP's. Repeat Loop until no new GWPs are added to GWP-New. 
   \item[\emph{Step E}:] Repeat Step C, until all GWPs have been added to new GWPs.  
   \item[\emph{Step F}:] Return to Step C for next PES, continuing until all PES are checked.
   \item[\emph{Step G}:] Once all new, consolidated GWPs are generated, calculate average weight of GWPs.  For all GWPs, if the weight is less than a threshold percentage of the average weight ( we have used 3\%) then discard the GWP from further propagation.
   \item[\emph{Step H}:] Procedure (starting from Step B) can be repeated, until no further change in the GWPs occurs, to attempt further consolidation. 
   \item[\emph{Step I}:] For all GWPs on different PES: If the overlap of the normalized GWPs is greater than a threshold, they are re-connected and undergo Coupled Propagation (Step 4), else if no ``partner" is found then the GWP undergoes independent propagation (Step 2). 
\end{itemize}
 \item [Monte-Carlo Sampling of many branch GWPs (Optional Consideration)] \hfill \\
 If a GWP is the result of many branches one expects its contribution to be small, with many similar GWPs contributing. Thus for branches of higher order than a defined number $B_{max}$, one may choose to sample the branching by Monte-Carlo rather than explicitly propagating both branches.
  \begin{itemize}
  \item[\emph{Step 5/6-a}:] If connection between GWPs is ended, and the GWPs have branched more than $B_{max}$ times, choose one trajectory to propagate by random number generation (Monte Carlo), based on the weights of the GWP. The propagated GWP will have new weight which is equal to the sum of the weights of the two GWPs.
   \item[\emph{Step 5/6-b}:] New GWPs generated by the Consolidation and Filtering procedure will be considered as having branched zero times. 
\end{itemize}
 \item [Miscellaneous considerations to limit branches (Optional Considerations)] \hfill \\
  \begin{itemize}
  \item[\emph{6-M}:]  (Modified Step 6) If GWP leave the region of significant non-adiabatic coupling (NAC), ${\bf d}_{n,m}^T \frac{{\hat m}^{-1} \{{\bf p}_0+{\bf p}_1\}/2}{\vert {\hat m}^{-1} \{{\bf p}_0+{\bf p}_1\}/2 \vert} < D_{min}$, eliminate connection in single direction, allowing GWP which is still inside NAC region to add to the GWP which has left, but not the reverse. This can help prevent the generation of many small GWPs as the coupled GWPs leave the region of NAC at slightly different times. 
   \item[\emph{Step CP-M}:]  (Modification to Coupled Propagation) Only check whether to break connection after GWPs are connected for a  finite time ($t_{min}= 10$ a.u.). This helps prevent generation of many small wave packets in difficult, highly-chaoitc, regions or near edges of NAC region. 
   \item[\emph{Step 3-M}:] (Modification to Step 3) One can define a region of significant NAC where the Massey parameter\cite{Massey} for each GWP:
   \begin{align}
   \zeta_0= \vert \frac{{\bf d}_{n,m}^T {\hat m}^{-1} {\bf p}_0}{E_n-E_m} \vert ~/~\zeta_1= \vert \frac{{\bf d}_{m,n}^T {\hat m}^{-1} {\bf p}_1}{E_m-E_n} \vert >\zeta_{min}~.
   \end{align}
   Wwe use $\zeta_{min}=1E-3$. One may also limit the NAC region to areas where $\frac{{\bf d}_{n,m}^T \{{\bf p}_0-{\bf p}_1\}}{{\bf d}_{n,m}^T \{{\bf p}_0+{\bf p}_1\} } < \Delta_{max}$ (we use  $\Delta_{max}=1.5$). When this ratio is large rapidly oscillating phase differences between generated GWPs is expected to cancel out. 
   \item[Note:] With all the optional considerations, there is a threshold parameter $(O_{min}, D_{min}, B_{max},$ $t_{min},\zeta_{min}, \Delta_{max})$ which can be used to tune accuracy vs efficiency. Convergence with the thresholds can be checked to determine if the information loss is acceptable. 
\end{itemize}
\end{description}
\section{Non-orthogonality of Basis}
During Coupled Propagation, we must take a superposition of two Gaussians which are in non-orthogonal basis states. For two coupled GWPs, which generate two ``hopped" GWPs we have:
\begin{align}
\label{NO}
\vert \Psi \rangle = g_1 \vert 1 [{\bf x}_1] \rangle + g_{1,2} \vert 2 [{\bf x}_1] \rangle + g_{2,1} \vert 1 [{\bf x}_2] \rangle + g_2 \vert 2 [{\bf x}_2] \rangle
\\
\nonumber
\approx  G_1  \vert 1 [{\bf x}_1] \rangle  + G_2  \vert 2 [{\bf x}_2] \rangle  \,.
\end{align}
Here, $g_{1(2)}$ is the GWP initially on PES 1 (2) which stays on PES 1(2), and $g_{1,2 (2,1)}$ is the GWP which is initially on PES 1 (2) and hops to PES 2(1). The GWPs $g_{1,2}$ and $g_{2,1}$ must be rotated from the electronic states $  \vert 2 [{\bf x}_1] \rangle$ and $\vert 1 [{\bf x}_2] \rangle$ to  $\vert 2 [{\bf x}_2] \rangle$ and $\vert 1 [{\bf x}_1] \rangle $ respectively. Projection of $  \vert 2 [{\bf x}_1] \rangle$ onto the orthogonal basis  $\vert 2 [{\bf x}_2] \rangle + \vert 1 [{\bf x}_2] \rangle$ will result in the wavepacket being ``duplicated":
\begin{align}
g_{1,2} \vert 2 [{\bf x}_1] \rangle = g_{1,2} \, \text{cos} \theta \, \vert 2 [{\bf x}_2] \rangle + g_{1,2} \, \text{sin} \theta \, \vert 1 [{\bf x}_2] \rangle~,
\end{align} 
and similarly, we have:
\begin{align}
g_{2,1} \vert 1 [{\bf x}_2] \rangle = g_{2,1} \, \text{cos} \theta \, \vert 1 [{\bf x}_1] \rangle + g_{2,1} \, \text{sin} \theta \, \vert 2 [{\bf x}_1] \rangle~.
\end{align} 
where $\text{cos} \theta \equiv \langle 1 [{\bf x}_2]  \vert 1 [{\bf x}_1] \rangle \equiv \langle 2 [{\bf x}_2]  \vert 2 [{\bf x}_1] \rangle$ and $\text{sin} \theta \equiv \langle 1 [{\bf x}_2]  \vert 2 [{\bf x}_1] \rangle \equiv \langle 2 [{\bf x}_1]  \vert 1 [{\bf x}_2] \rangle$.
This step can be repeated, and since by definition $\vert \text{sin}\theta \vert <1$ the infinite series of rotations will converge. Thus Equation \ref{NO} can be re-expressed as:
\begin{align}
\vert \Psi \rangle = g_1 \vert 1 [{\bf x}_1] \rangle + g_{1,2} \vert 2 [{\bf x}_1] \rangle + g_{2,1} \vert 1 [{\bf x}_2] \rangle + g_2 \vert 2 [{\bf x}_2] \rangle
\\
\nonumber
= \Big[ g_1 + \frac{1}{\text{cos} \theta}g_{2,1} + \frac{\text{sin} \theta}{\text{cos} \theta} g_{1,2} \Big] \vert 1 [{\bf x}_1] \rangle   + \Big[ g_2 + \frac{1}{\text{cos} \theta}g_{1,2} + \frac{\text{sin} \theta}{\text{cos} \theta} g_{2,1} \Big] \vert 2 [{\bf x}_2] \rangle
\\
\nonumber 
\approx  G_1  \vert 1 [{\bf x}_1] \rangle  + G_2  \vert 2 [{\bf x}_2] \rangle  \,.
\end{align}

Thus, not only does the ``hopped" GWP contribute to the reconstructed GWP on the final surface, but it also has a contribution to the re-constructed GWP on the same surface. However this contributions is small for shifts in position ${\bf x}_1-{\bf x}_2$ in which the Coupled GWP dynamics works $(\text{cos} \theta \approx 1)$ . That is, GWPs should branch before this non-orthogonal basis effect becomes relevant. In the calculations presented in the paper, inclusion of this effect does not change results significantly. We present it here for formal completeness, and for possible future importance. 
 
\bibliographystyle{apsrev4-1}